\documentclass[aps,prl,superscriptaddress,twocolumn,showpacs,floatfix,a4paper,reprint]{revtex4-1}

\usepackage{amsfonts,amsmath,amssymb}
\usepackage{textcomp}
\usepackage{bm}
\usepackage{upgreek}
\usepackage{graphicx}
\usepackage{natbib}
\usepackage{color}
 \usepackage{times}
  \usepackage{bm}
\usepackage[normalem]{ulem}
\newcommand{\ie}{\textit{i.e.}}

\renewcommand{\vec}[1]{\bm{#1}}
\newcommand{\ee}{\mathrm{e}}
\newcommand{\ii}{\mathrm{i}}

\newcommand{\Tr}{\mathrm{Tr}}

\newcommand{\nablabf}{\boldsymbol{\nabla}}

\newcommand{\rot}{\nablabf\times}

\newcommand{\AAA}{\vec{A}}

\newcommand{\KKK}{\vec{K}}

\newcommand{\qqq}{\vec{q}}

\newcommand{\rrr}{\vec{r}}

\newcommand{\uuu}{\vec{u}}

\newcommand{\eps}{\epsilon}

\newcommand{\bra}[1]{\langle #1 |}
\newcommand{\ket}[1]{|#1\rangle}

\newcommand{\bsub}{\begin{subequations}}
	\newcommand{\esub}{\end{subequations}}
\def\bal#1\eal{\begin{align}#1\end{align}}
\def\bsubal#1\esubal{\bsub \begin{align}#1\end{align} \esub}

\renewcommand{\eqref}[1]{Eq.~(\ref{eq:#1})}

\usepackage[normalem]{ulem}

\begin{document}
   	
\title{Valley-Polarized Quantum Transport Generated by Gauge Fields in Graphene}	

\author{Mikkel Settnes}\email[]{mikse@nanotech.dtu.dk}
\affiliation{Center for Nanostructured Graphene (CNG), Department of Micro- and Nanotechnology Engineering, Technical University of Denmark, DK-2800 Kgs. Lyngby, Denmark}
\affiliation{Department of Photonics Engineering, Technical University of Denmark, DK-2800 Kgs. Lyngby, Denmark}
\author{Jose H. Garcia}
\affiliation{Catalan Institute of Nanoscience and Nanotechnology (ICN2), CSIC and The Barcelona Institute of Science and Technology, Campus UAB, Bellaterra, 08193 Barcelona, Spain}
\author{Stephan Roche}
\affiliation{Catalan Institute of Nanoscience and Nanotechnology (ICN2), CSIC and The Barcelona Institute of Science and Technology, Campus UAB, Bellaterra, 08193 Barcelona, Spain}
\affiliation{ICREA - Institucio Catalana de Recerca i Estudis Avancats, 08010 Barcelona, Spain}
 
\date{\today}
 
%

%
%



\begin{abstract}
We report on the possibility to simultaneously generate in gaphene a {\it bulk valley-polarized dissipative transport} and a {\it quantum valley Hall effect} by combining strain-induced gauge fields and real magnetic fields. Such unique phenomenon results from a ``resonance/anti-resonance" effect driven by the superposition/cancellation of superimposed gauge fields which differently affect time reversal symmetry. The onset of a valley-polarized Hall current concomitant to a dissipative valley-polarized current flow in the opposite valley is revealed by a $e^2/h$ Hall conductivity plateau. We employ efficient linear scaling Kubo transport methods combined with a valley projection scheme to access valley-dependent conductivities and show that the results are robust against disorder.\end{abstract}
\maketitle
  
In graphene and other two-dimensional materials, degenerate valleys of energy bands, well-separated in momentum space, constitute a discrete degree of freedom for low-energy carriers, provided intervalley mixing is negligible (case of long range disorder). Such valley degree of freedom can then be seen as a non-volatile information carrier, provided that it can be coupled to external probes. Similarly to research in graphene spintronics \cite{Han2014_NATNANO,ROC_2DM3}, valleytronics and controlling the valley degree of freedom in graphene and other 2D materials has attracted a considerable attention \cite{Lee2016_NAT,Lensky2015_PRL,Shimazaki2015_NATPHY}, and many studies have investigated different options to realize valley polarized current or to filter electrons with a given valley polarization \cite{RYC_NP3, XIA_PRL99,ZHA_APL93, WAK_IJMPB16,GUN_PRL106,Zhenhua2011_PRL,Settnes_2016_PRL}. In presence of broken inversion symmetry, the valley index is also predicted to play a similar role as the spin degree of freedom in phenomena such as Hall transport, magnetization, optical transition selection rules, and chiral edge modes \cite{LU_PRB81, YAO_PRL102, CHE_PRB77, LIU_AP59}.  Recently, it has been shown that for modest levels of strain, graphene can also sustain a classical valley Hall effect that can be detected in nonlocal transport measurements \cite{Zhang2017}. All this stimulates the search for efficient control of valley dynamics by magnetic, electric, and optical means, which would form the basis of valley based information processing. 

On the other hand,  it was soon realized that non-uniformly strained graphene can be modeled by the inclusion of a gauge field in the effective Hamiltonian (though it is not Berry-like). Such gauge field preserves time reversal symmetry, but induces pseudomagnetic fields (PMFs) of opposite signs in the two valleys forming the low-energy electronic bandstructure \cite{Fujita2011_JAP}. In particular, mechanical deformations aligned along three main crystallographic directions were predicted to generate strong gauge fields, acting effectively as an uniform magnetic field, opening the possibility to trigger a pseudomagnetic quantum Hall effect for strained superlattices \cite{Guinea_2009_NATPHY}. Experimentally, evidences of PMFs with values varying from few tens up to hundreds of Tesla have been reported \cite{Levy_2010_SCI,Lu_2012_2012}. Finally, Gorbachev and coworkers recently measured intriguingly large nonlocal resistance signals in a situation where the alignment of a graphene monolayer onto a h-BN substrate results in a bandgap opening \cite{GOR_SCI346}. Their interpretation proposes that the formation of {\it bulk topological valley currents} is intertwined with the presence of a Berry curvature generated by the mass term \cite{Song2015}, a scenario which is under questioning \cite{KIR_PRB92,Cresti_2016_RNC}. 
Accordingly to date, despite the wealth of theoretical proposals of valley dependent effects  \cite{FUJ_APL97,ZHA_PRB82,ZHA_JPCM23,SON_APL103,MYO_CAP14,GAR_PRL100,PER_JPCM21,HSI_APL108,ABE_APL95,COS_PRB92,PET_NJP14,PRA_APL104,Milo_2016_APL,Carrillo-Bastos_2016_PRB,Jiang_2013_PRL}, experimental fingerprints of PMF on quantum transport and unambiguous demonstration of a valley Hall effect in graphene remain elusive. 

Here, we predict that once the electronic structure of Dirac fermions embeds a strain-related gauge field, it is possible to fine-tune the superposition of an external real magnetic field to reach a resonant effect, where the sum of valley-dependent effective magnetic fields either sum up or cancel each other. This results in a remarkable valley-polarized quantum transport regime, exhibiting simultaneously a quantum valley Hall effect (QVHE) and a valley-polarized dissipative transport regime. To achieve such results, we use a tight-binding framework together with an efficient linear scaling quantum transport methodology and a novel valley projection scheme which gives access to valley-dependent quantum transport quantities, even in presence of disorder. 

\textit{Methodology.---} We employ a first nearest neighbor tight-binding (TB) model where strain is included through a modification of the hopping parameters while the external magnetic field is added using a standard Peierls substitution \cite{Pereira_2009_PRB}.  In the Dirac approximation the strain is described by a gauge field $\pm \AAA_{\rm S}$, where $\pm$ denote the two valleys. Such gauge field is related to the strain tensor $\eps_{ij}$ through $\AAA_{\rm S} \propto \big(\eps_{xx} -\eps_{yy},-2\eps_{xy}\big)$ \cite{Guinea_2009_NATPHY,Vozmediano2010_PR,Fujita2011_JAP}, and the pseudomagnetic field becomes $B_{\rm S} = \rot \AAA_{\rm S}$. From this, it is straightforward to show that a triaxial deformation $\uuu(x,y)=u_0 \big(2xy,x^2-y^2\big)$ induces a constant PMF. Uniaxial tensile strain have also been shown to generate constant PMF \cite{Zhu_2015_PRL}.
In the presented calculations, we use a 100 nm $\times$ 100 nm graphene sample ($\sim 4\times10^5$ atoms) with a maximum strain of $\Delta_m \approx 8 \%$ corresponding to a PMF of 50 T. The maximum strain is obtained along the edge of the sample, so all results can be rescaled, and for instance keeping $\Delta_m =8$ \%, we find a PMF of 5 T for a sample of approximately 1 $\mu$m. 
The sample choice also implies that not all part of the sample experience a uniform PMF. This happens along the edge of the samples where non-uniform PMF will act as scattering centers that can mix valleys. The presented results are robust against this type of valley mixing as we considering a bulk effect happening in the part of the sample with a constant PMF. The results remain qualitatively unchanged as long as a sufficiently large part of the sample experience a uniform field.
 	
The valley related physics can be investigated either using the Dirac's equation  \cite{AND_JPSJ84,Cresti_2016_RNC,Cazalilla2D2017} or by performing non-local transport calculations using the Landauer-B\"uttiker formalism in Hall bar geometries \cite{KIR_PRB92,Cresti_2016_RNC}. The former can be cumbersome for arbitrary types of disorder, while the later becomes computationally prohibitive for large scale systems. Here, the use of a real space TB method gives access to simulations of very large scale disorder systems (with up to several ten millions of atoms), accounting for disorder-induced valley mixing and other distortions of the Fermi surface \cite{Manes2013_PRB,RamezaniMasir2013_SSC,Pellegrino2011_PRB}. To compute the valley conductivity tensor, we develop an efficient real-space quantum transport method using the Kubo-Bastin formula \cite{AND_JPSJ84,Bastin1971,Bruno2001_PRB}
 
\begin{align} 
\sigma_{\alpha \beta} & = \frac{\ii \ee^2 \hbar}{\Omega}\int_{-\infty}^{\infty}d\eps f(\eps) 
\Tr\big[  v_\alpha \delta(\eps-H) v_\beta \frac{dG^r(\eps)}{d\eps} \nonumber \\
 &- v_\alpha  \frac{dG^a(\eps)}{d\eps} v_\beta\delta(\eps-H)\big], \label{Kubo}
\end{align}
where $\Omega$ is the area of the sample, $f(\eps)$ is the Fermi-Dirac distribution, $v_{\alpha}$ is the $\alpha$ component of the velocity and $G^{r/a}(\eps)=(\eps - H \pm \ii \eta)^{-1} $ 
are the advanced/retarded Green's functions. The Kubo-Bastin formula can be computed efficiently by expanding  $\delta(\eps-H)$ and $G^{r/a}(\eps)=(\eps - H \pm \ii \eta)^{-1}$ in terms of Chebyshev polynomials \cite{Silver_1997_PRE,Weisse_2006_RMP,Garcia_2015_PRL,Cresti_2016_RNC}. As shown in Ref. \cite{Garcia_2015_PRL}, the conductivity becomes
\begin{align} 
\sigma_{\alpha\beta} &= \frac{4\ee^2 \hbar}{\pi\Omega} \frac{4}{\Delta E^2}\int_{-1}^{1}d\tilde{\eps} \frac{f(\tilde{\eps})}{(1-\tilde{\eps}^2)^2} \sum_{m,n}^M \Gamma_{nm}(\tilde{\eps}) \mu_{nm}^{\alpha \beta}, \label{sigma_cheb}
\end{align} 
where $\Delta E = E_{\rm max} - E_{\rm min}$ is the range of the energy spectrum, $\Gamma_{mn}(\tilde{\eps})$ is an energy dependent function  given by $(\tilde{\eps}-\ii n \sqrt{1-\tilde{\eps}^2})\ee^{\ii n \arccos(\tilde{\eps})} T_m(\tilde{\eps})$ \ $+ (\tilde{\eps}+\ii m \sqrt{1-\tilde{\eps}^2})\ee^{-\ii m \arccos(\tilde{\eps})}T_n(\tilde{\eps})$, and $\mu_{nm}^{\alpha \beta}$ are the Chebyshev expansion moments 
\begin{align}
\mu_{nm}^{\alpha \beta}  &= \frac{g_ng_m}{(1+\delta_{n,0})(1+\delta_{m,0})}  \Tr\big[T_n(\tilde{H}) v_\alpha T_m(\tilde{H}) v_\beta \big].\label{moment}
\end{align}
Here $T_n(x)$ are the Chebyshev polynomials defined through the recurrence relation $T_{n+1} (x) = 2xT_{n}(x) - T_{n-1}(x)$, with $T_0(x)=1$ and $T_1(x) = x$. $\tilde{H}$ and $\tilde{\varepsilon}$ are the rescaled Hamiltonian and energies so that the energy spectrum  lies within the interval $[-1,1]$.The factor $g_n$ is known as the Jackson kernel and is incorporated to dampen the Gibbs oscillations emerging from the truncation of the expansion in Eq. (\ref{sigma_cheb}) \cite{Weisse_2006_RMP}. The trace in Eq. (\ref{moment}), is efficiently calculated using the stochastic average scheme presented in \cite{Weisse_2006_RMP,Drabold_1993_PRL,Iitaka2004_PRE}.
 
The expression in Eq. (\ref{sigma_cheb}) has been used for computing the spin Hall conductivity \cite{Garcia_2016_2D,Cresti_2016_RNC} but is more general and can be applied to any type of current induced by an external electric field. We here extend the use of Eq. (\ref{sigma_cheb}) to compute the valley conductivity tensor by introducing the valley-velocity operator $v_{\beta}^v$ defined as
\begin{equation}
v_{\beta}^{v}\equiv P_{\KKK}v_\beta P_{\KKK}  - P_{\KKK'}v_\beta P_{\KKK'}\label{ValleyDef}
\end{equation}
where  $P_{\KKK}(P_{\KKK'})$ is the valley projection operator onto the $\KKK$($\KKK'$) valley, defined as  $P_{\KKK} = \sum_{\qqq} w_{\qqq} \ket{\qqq}\bra{\qqq}$ where the sum runs over the Brillouin zone. The weight $w_{\qqq}$, enforcing a projection to a single valley, is given by $w_{\qqq} = \Theta(|\qqq-\KKK|<q_c)$, where $q_c$ is a cutoff that can be conveniently chosen between $\KKK$ and $\KKK'$ e.g. $q_c=\pi/(3\sqrt{3}a_0)$. Applying the operator $P_{\KKK}$ to a real space vector corresponds to Fourier transforming it on a Monkhorst-Pack k-grid while keeping only coefficients satisfying the condition $|\qqq-\KKK|<q_c$. To use the resulting vector in a real space basis, the inverse Fourier transfor is further applied. This operation is $\mathcal{O}(N\log(N))$, which is much more efficient than the $\mathcal{O}(N^3)$ exact evaluation of the non-sparse matrix $P_{\bm{K}}$.
Finally, the valley conductivity tensor is obtained by modifying the Chebyshev expansion moments $\mu_{nm}^{\alpha \beta} \rightarrow \mu_{nm}^{\alpha \beta(v)}$ , where
\begin{equation} 
\mu_{nm}^{\alpha \beta(v)} = \frac{g_ng_m}{(1+\delta_{n,0})(1+\delta_{m,0})}  \Tr\big[T_n(\tilde{H}) v_\alpha T_m(\tilde{H}) v_\beta^v \big].\label{Valleymoment}
\end{equation}
This definition of the valley-velocity operator in a system with decoupled valleys, matches the definition given in Ref. \cite{AND_JPSJ84} in the continuous limit, and takes into account all the contributions of intravalley and intervalley scattering to the conductivity through the full Hamiltonian. 

 \begin{figure}[tb!]
 	\begin{center}
 		\includegraphics[width= 0.95\columnwidth]{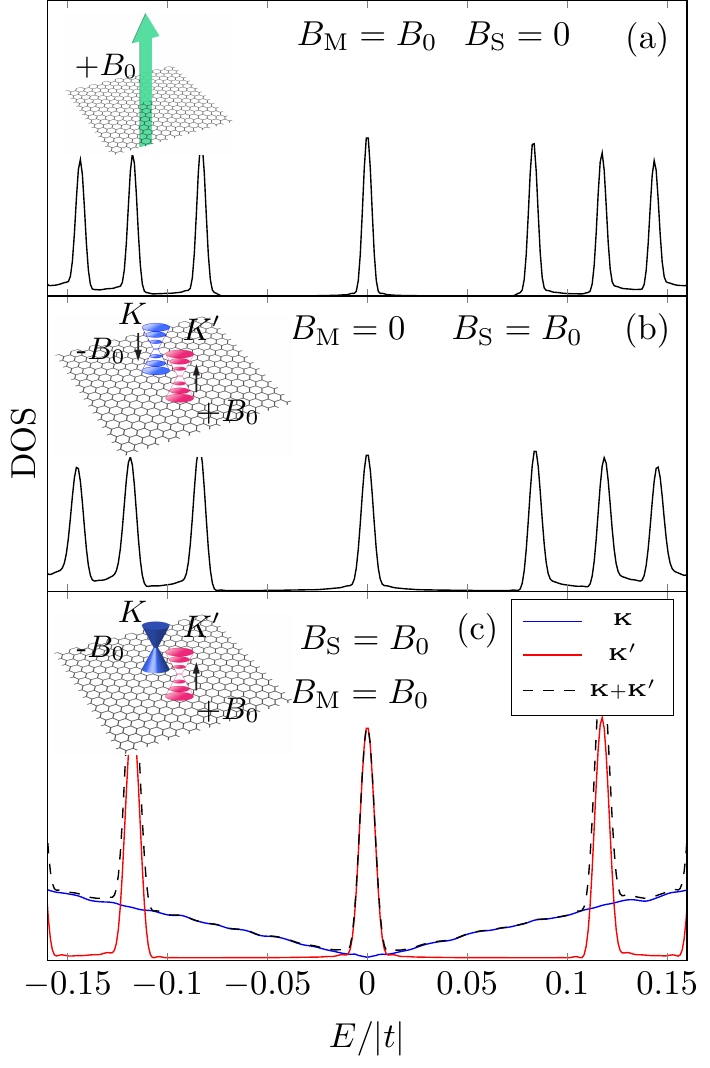}
 		\caption[]{DoS of graphene in presence of (a) an external magnetic field $B_{\rm M} = B_0$, (b) a strain-induced PMF $B_{\rm S}=B_0$. (c) valley polarized DoS for graphene with both a strain induced PMF and external magnetic field with $B_{\rm S} = B_{\rm M} = B_0$. We take $B_0 = 50$ T. We use 200 random vectors and 4000 moments which set the energy resolution in $\approx$ 6 meV \cite{Weisse_2006_RMP}. } \label{fig_dos}
 	\end{center}
 \end{figure}
 
\textit{Density of states.---} 
Fig. \ref{fig_dos} gives the density of states (DoS) for three different cases. The presence of an external magnetic field is described by a gauge vector potential $\AAA_{\rm M}$ identical for the two valleys and producing the standard Landau quantization in graphene \cite{Goerbig2011} (Fig. \ref{fig_dos}(a)).  The DoS for the strained graphene structure, in absence of external field is shown in Fig. \ref{fig_dos}(b). Here the PMF (related to $\AAA_{\rm S}$) has opposite orientation in the two valleys, ensuring the overall time-reversal symmetry \cite{Vozmediano2010_PR,LOW_NL10,Qi_2013_NL}. 
At last, we combine both an external magnetic field and a strain-induced PMF. We choose the external magnetic field to be equal to the PMF, $B_{\rm S}=B_{\rm M}=B_0$. This induces a valley anisotropy since the effective vector potential becomes $\AAA_{\rm M}\pm \AAA_{\rm S}$, depending on the valley. In Fig. \ref{fig_dos}c, the Landau quantization in valley $K'$ corresponds to a B-field of $2B_0$, instead of $B_0$ as expected \cite{MorpurgoGuineaPRL2006}. The DoS in the opposite valley is half that of pristine graphene. Since the PMF also gives rise to a pronounced sublattice polarization \cite{Settnes_2015_PRB,Schneider_2015_PRB,Neek-Amal_2013_PRB,Settnes_2016_PRB,Georgi2016_ArX}, a special state is created for the zeroth Landau level. This state is both valley and sublattice polarized, meaning that only one sublattice and one valley contributes to the density of states. 

\begin{figure}[tb!]
	\begin{center}
		\includegraphics[width= 0.95\columnwidth]{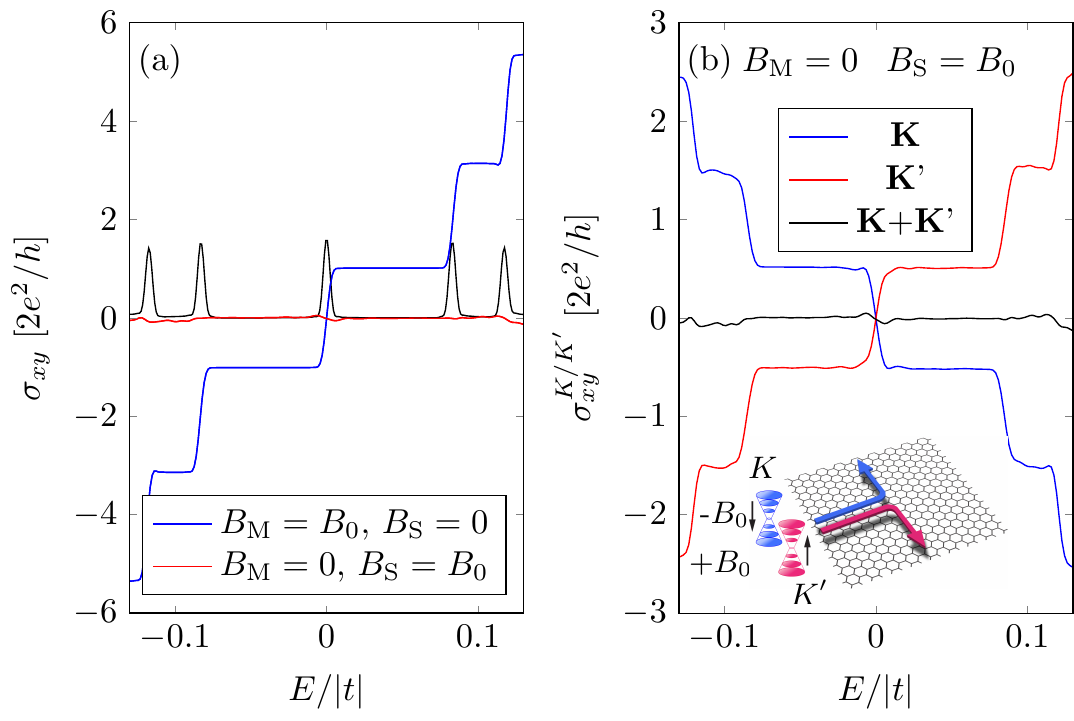}
		\caption[]{(a) Hall conductivity in the presence of an external magnetic field (blue) and a strain field inducing a PMF (red). The DoS for real magnetic field is indicated for reference (black). 
		(b)  $\sigma_{xy}^K$ (blue) and $\sigma_{xy}^{K'}$ (red) showing the QVHE.We use 200 random vectors and 4000 moments which set the energy resolution in $\approx$ 6 meV \cite{Weisse_2006_RMP}. 
		Inset: Illustration of the spatial splitting of both valleys.
		} \label{fig_Hall1}
	\end{center}
\end{figure}
 
\textit{Conductivity and valley Hall effect.---} The Hall conductivity $\sigma_{xy}$ in the presence of an external magnetic field (blue curve) and a strain-induced PMF (red curve) is analyzed in Fig. \ref{fig_Hall1}a. The external magnetic field gives rise to the quantum Hall effect exhibiting the standard $2e^2/h$ plateau. Differently, $\sigma_{xy}$ for the strained graphene reveals a remarkable difference since the PMF does not give rise to any Hall signal, \ie \; $\sigma_{xy}\approx 0$. However, the valley resolved calculation (Fig. \ref{fig_Hall1}b), shows that even though the PMF keeps the overall time reversal symmetry, it breaks the time reversal invariance in the individual valleys. Consequently, the PMF gives rise to a QVHE with opposite sign of $\sigma_{xy}^{K/K'}$ in the two valleys such that the total charge Hall current $\sigma_{xy}=\sigma_{xy}^{K}+\sigma_{xy}^{K'}$ is zero. In conclusion the electronic system does not experience a zero B-field {\it charge} Hall effect, but instead a quantum \textit{valley} Hall effect reminiscent of the quantum spin Hall effect \cite{Bernevig_2006_PRL,Kane_2005_PRL}. The Hall plateaus of height $e^2/h$ are halved compared to the usual quantum Hall regime (see Fig. \ref{fig_Hall1}a) because only one valley is involved. Such QVHE originates from the splitting of an incoming current into the two different valleys originally moving in the same direction (see inset of Fig. \ref{fig_Hall1}b). This spatial valley splitting could be explored experimentally in non-local transport measurements. Indeed, this situation, first proposed in \cite{Guinea_2009_NATPHY} appears similar, although of different origin, to the theoretical picture introduced in Ref. \cite{GOR_SCI346,Song2015}. 

\begin{figure}[tb!]
 	\begin{center}
 		\includegraphics[width= 0.95\columnwidth]{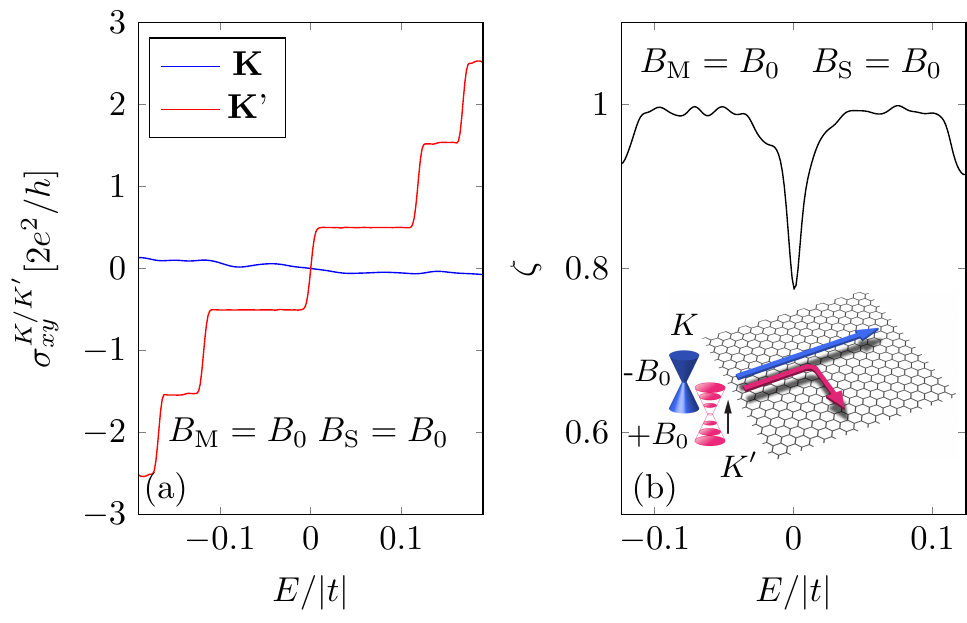} 
 		\caption[]{(a) Valley polarized $\sigma_{xy}$. 
 		(b) Valley polarization $\zeta = (\sigma_{xx}^{K}-\sigma_{xx}^{K'})/\sigma_{xx}$ of the dissipative longitudinal current. Inset: Illustration of valley Hall effect in $K'$ and standard diffusive transport for $K$ in the longitudinal direction.} \label{fig_Hall2}
 	\end{center}
 \end{figure}

Here, to obtain a non-zero charge Hall signal, $\sigma_{xy} = \sigma_{xy}^K+\sigma_{xy}^{K'}$, we apply an external magnetic field to the strained graphene sample. When both gauge fields match in amplitude, they cancel in one valley recovering the clean Dirac cone. The opposite valley gets a doubling of the effective magnetic strength, although the superimposed gauge fields have completely different origin. Fig. \ref{fig_Hall2} shows the formation of the QVHE for the valley with enhanced gauge field, with quantization steps of $e^2/h$ for $\sigma_{xy}^{K'}$, whereas $\sigma_{xy}^{K}=0$ exhibits no Hall signal. Effectively only one valley is deflected in this scenario, yielding a valley polarized charge Hall signal. In the "pristine" valley, we get a normal diffusive current in the longitudinal direction. Consequently, the quantized valley-polarized Hall current is accompanied by a valley polarized diffusive transport in the longitudinal direction for energies in-between the Landau quantization (see inset of Fig. \ref{fig_Hall2}b). The valley polarization is $\zeta = (\sigma_{xx}^{K}-\sigma_{xx}^{K'} )/ (\sigma_{xx}^{K}+\sigma_{xx}^{K'} ) \approx 1$ for the longitudinal current at energies in-between the Landau levels, but is reduced at the Landau levels where both valleys has a non-zero longitudinal contribution (see Fig. \ref{fig_Hall2}b).

  \begin{figure}[tb!]
  	\begin{center}
  		\includegraphics[width= 0.95\columnwidth]{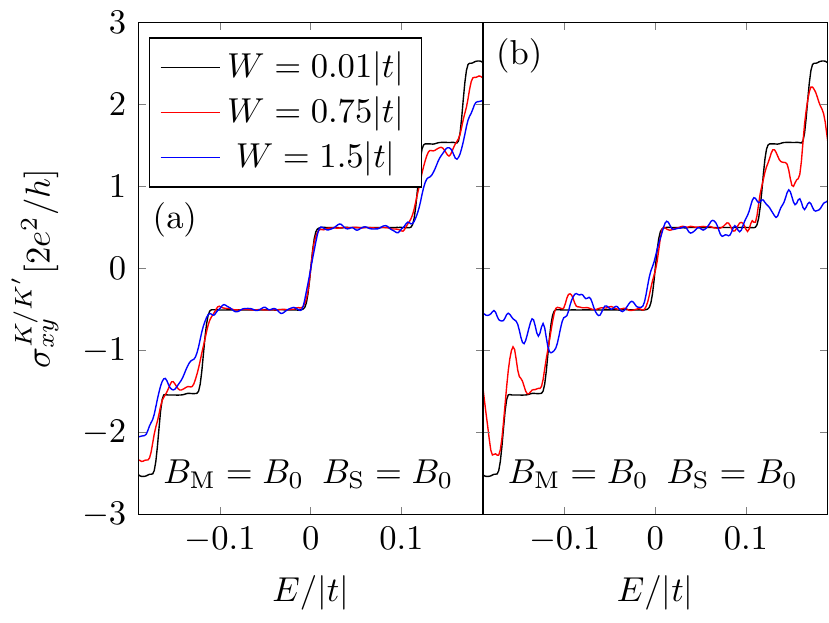}   
  		\caption[]{$\sigma_{xy}^{K'}$ in the presence of both external magnetic field and deformation field (pseudomagnetic field) for different amount of (a) long range and (b) short range disorder. }
\label{fig_disorder}
  \end{center}
  \end{figure}
  
\textit{Disorder.---} Finally, we study how disorder affects the valley polarized quantum transport regime. Short range Anderson disorder is described by onsite energies $\eps_i$ chosen randomly within $[-W/2,W/2]$, whereas long range disorder is modelled by Gaussian impurities at positions $\rrr_n$ with a concentration of 0.1, $V_n = \sum_i \eps_i \exp\big(-|\rrr_n-\rrr_i|^2/(2\xi^2)\big)$ with $\eps_i \in [-W/2,W/2]$ and $\xi=5a_0$ \cite{Roche_2012_SSC}. Fig. \ref{fig_disorder} shows $\sigma_{xy}^{K'}$ in the presence of both gauge fields and types of disorder. Long range disorder, which is not expected to induce significant valley mixing, only causes minor perturbations to the QVHE associated with the strain-induced PMF. Conversely, short range disorder is expected to produce large intervalley scattering \cite{CastroNeto2009_RMP}, which should strongly damage the quantization of Hall plateaus. However, surprisingly, the first plateau remains quite robust against a significant amount of disorder, whereas the higher order plateaus disappear for smaller disorder strengths. This is a manifestation of the special nature of the $n=0$ pseudo-Landau level \cite{Neek-Amal_2013_PRB,Settnes_2016_PRB,Settnes_2016_2D}. It experiences a complete sublattice polarization with both valleys localized on one sublattice, meaning that perturbations made to one sublattice does not provoke the same valley mixing effect as perturbations made to the other. Consequently, the special nature of the $n=0$ pseudo-Landau level makes the QVHE even more robust to valley mixing.

\textit{Conclusion.---} The combination of strain-induced gauge and external magnetic fields results in a remarkable ``resonance" phenomenon in which the total gauge disappears in one valley and doubles in the other. This is reflected in the charge Hall signal which becomes fully valley-polarized with a $e^2/h$ plateau instead of the standard $2e^2/h$. Simultaneously, the dissipative transport coefficient is dominated by a valley-polarized contribution. The results therefore offer an experimental smoking gun for a fully realized valley-polarized quantum transport regime.
  
\textbf{Acknowledgements:} The work of M.S is supported by the Danish Council for Independent Research (DFF-5051-00011). The Center for Nanostructured Graphene (CNG) is sponsored by the Danish National Research Foundation (DNRF103). This work has been performed thanks to the computational resources awarded from PRACE and the Barcelona Supercomputing Center (Mare Nostrum), under Project No. 2015133194. This work has also received funding from the European Union Seventh Framework Programme under grant agreement 604391 Graphene Flagship. S.R. acknowledges the European Union Seventh Framework Programme under grant agreement 604391 Graphene Flagship, the Spanish Ministry of Economy and Competitiveness for funding (MAT2012-33911), the Secretaria de Universidades e Investigacion del Departamento de Economia y Conocimiento de la Generalidad de Cataluna and the Severo Ochoa Program (MINECO SEV-2013-0295).

\end{document}